\documentclass[]{spie}  

\usepackage{amsmath,amsfonts,amssymb}
\usepackage{graphicx}
\usepackage[colorlinks=true, allcolors=blue]{hyperref}
\usepackage{cleveref}
\usepackage{float}
\usepackage{tikz}

\title{CCAT: Multi-rate DSP for sub-mm astronomy: polyphase synthesis filter bank on FPGA for enhanced MKID readout}

\author[a,*]{Ruixuan(Matt) Xie}
\author[b]{Adrian K. Sinclair}
\author[b]{James Burgoyne}
\author[b,c]{Scott Chapman}
\author[d]{Anthony Huber}

\affil[a]{Dept. of Electrical and Computer Engineering, University of British Columbia, Vancouver, British Columbia, Canada}
\affil[b]{Dept. of Physics and Astronomy, University of British Columbia, Vancouver, British Columbia, Canada}
\affil[c]{Dept. of Physics and Atmospheric Science, Dalhousie University, Halifax, Nova Scotia, Canada}
\affil[d]{Dept. of Physics and Astronomy, University of Victoria, Victoria, British Columbia, Canada}

\begin{document} 
\maketitle
\footnotetext{This work has been published in Millimeter, Submillimeter, and Far-Infrared Detectors and Instrumentation for Astronomy XII, edited by Jonas Zmuidzinas, Jian-Rong Gao, Proc. of SPIE Vol. 13102, 1310213(2024). DOI: \href{https://doi.org/10.1117/12.3020586}{https://doi.org/10.1117/12.3020586}.}
\begin{abstract}
The next-generation mm/sub-mm/far-IR astronomy will in part be enabled by advanced digital signal processing (DSP) techniques. The Prime-Cam instrument of the Fred Young Submillimeter Telescope (FYST), featuring the largest array of submillimeter detectors to date, utilizes a novel overlap-channel polyphase synthesis filter bank (OC-PSB) for the AC biasing of detectors, implemented on a cutting-edge Xilinx Radio Frequency System on Chip (RFSoC). This design departs from traditional waveform look-up-table(LUT) in memory, allowing real-time, dynamic signal generation, enhancing usable bandwidth and dynamic range, and enabling microwave kinetic inductance detector (MKID) tracking for future readout systems. Results show that the OC-PSB upholds critical performance metrics such as signal-to-noise ratio (SNR) while offering additional benefits such as scalability. This paper will discuss DSP design, RFSoC implementation, and laboratory performance, demonstrating OC-PSB's potential in submillimeter-wave astronomy MKID readout systems.
\end{abstract}


{\noindent \footnotesize\textbf{*}Ruixuan(Matt) Xie, \href{mailto:Xieruix1@student.ubc.ca}{Xieruix1@student.ubc.ca}}

\section{INTRODUCTION}
\label{sec: intro}  
Far-infrared, broadly defined as wavelengths of 30 - 1000 µm, offers a rich and largely unexplored territory for discoveries. Observations in this band offer unique and significant value in astrophysics and cosmology, particularly in unveiling the origins of the universe. Recently in submillimeter (sub-mm) astronomy, microwave kinetic inductance detectors (MKIDs) have attracted popularity due to their high multiplexing factor and ease of manufacturing.\cite{farIF_review2019}

\begin{figure}[h]
    \centering
    \includegraphics[width=0.618\textwidth]{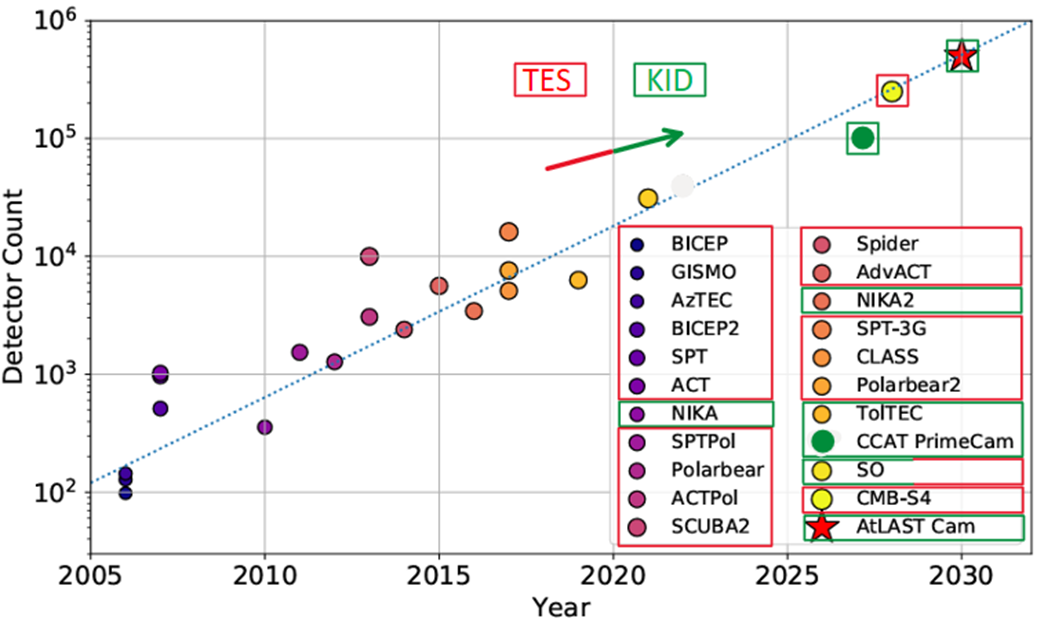}
    \caption{Detector type and count vs. year\cite{atlas_report} (annotated with the updated anticipated count of Prime-Cam detectors)}
    \label{fig: Increasing det count}
\end{figure}

The growing need for increased detector counts (as shown in \cref{fig: Increasing det count}\cite{atlas_report}) and the paradigm shift from time-domain multiplexing (TDM) to frequency-domain multiplexing (FDM) readout schemes have redefined the complexity bottleneck. Previously, the challenge lay in cryogenic wiring limitations. Today, the focus has shifted to warm-side back-end electronic signal processing capabilities. This evolution in detector technology necessitates advancements in readout electronics as well. ROACH2-based systems have given way to compact, high-speed, and low-power Radio Frequency System-on-Chips (RFSoCs).

Exemplifying the next step of MKID-based observations is the Fred Young Sub-mm Telescope (FYST), being developed by the CCAT collaboration\cite{CCAT_Prime_Collaboration_2022}. The Prime-Cam instrument, which will have the largest array of MKIDs to date, is the primary camera in FYST currently under development with first-light anticipated in 2026. The highest frequency 850-GHz instrument module\cite{Chapman850GHz2022} aboard Prime-Cam will have the most demanding detector count and bandwidth which are addressed by the overlap-channel polyphase synthesis filter
bank (OC-PSB) synthesizer designed and implemented in this study.

\begin{figure}[h]
    \centering
  \begin{minipage}[b]{0.3\linewidth}
    \includegraphics[width=\linewidth]{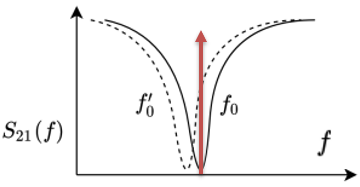}
    \caption{MKID transmission with probe/bias tone annotation \cite{sinclair2022techtalk}}
    \label{fig: SingleKIDreso}
  \end{minipage}
  \hspace{0.05\linewidth}
  \begin{minipage}[b]{0.4\linewidth}
    \includegraphics[width=\linewidth]{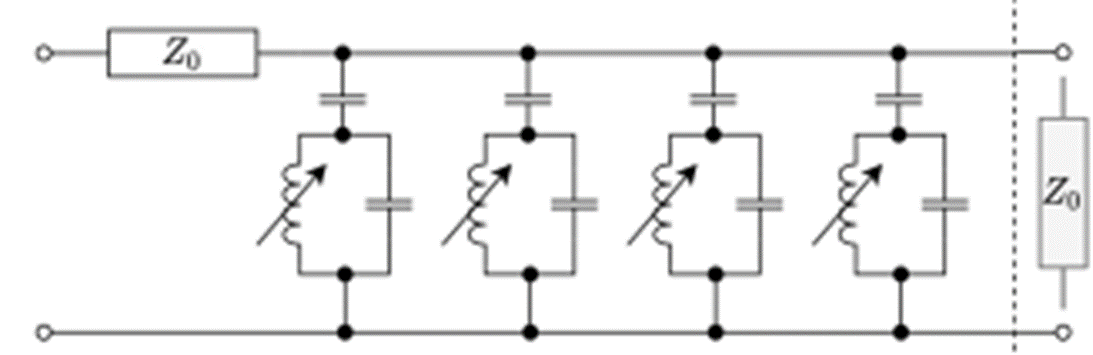}
    \caption{Simplified detector array equivalent circuit \cite{sinclair2022techtalk}}
    \label{fig: KIDsParallel}
  \end{minipage}
\end{figure}

MKIDs can be configured as microwave resonators with unique resonant frequencies as shown in \cref{fig: SingleKIDreso}. The resonant frequency is determined by a fixed capacitance, geometric inductance, and kinetic inductance. Thousands of resonators, in the form of an array, can be capacitively coupled to the transmission line as one RF network read by a single coaxial cable, as shown in \cref{fig: KIDsParallel}.

Photons are optically modulated before striking the detector array. Each MKID absorbs photon power and exhibits a change in $S_{21}$, the RF forward transmission, which can be simplified as a shift in resonant frequency for synthesizer design. The change in frequency response alters the probe tone power $P_{tone}$, frequency $f_{tone}$, and phase $\phi_{tone}$ observed by the receiving channelizer, thereby encoding the power received from the sky.

The goal of this study is to demonstrate the feasibility of employing an OC-PSB-based frequency synthesizer in an FPGA (Zynq UltraScale+ RFSoC ZCU111 Evaluation board) for the biasing of MKIDs, addressing the various challenges faced by the Prime-Cam instrument readout system while bringing enhanced capabilities. Specifically, the OC-PSB offers a novel approach to overcome the bandwidth (BW) and resolution limitations of traditional external memory look-up-table (LUT) methods\cite{sinclair2022ccatprime}, shifting the BW constrain from DDR4 memory to available FPGA resources. Synthesized inside the FPGA fabric, the tones can be edited in real-time, enabling enhanced readout capabilities such as ``tone tracking". Currently, the Prime-Cam instrument is aiming for a total detector count of  $\sim 75,000$, and $\sim 45,000$ detectors are from the 850-GHz instrument module. To achieve the desired mapping speed, the 850-GHz module is pursuing a readout of $\sim 1200$ detectors per radio-frequency (RF) network (read out by a single coaxial cable) across 1.024 GHz of bandwidth. \cite{TonySPIE2024}. The baseline readout\cite{sinclair2022ccatprime} bandwidth is constrained to 512 MHz by the DDR4 memory depth and data transfer rate. The OC-PSB will lift these constraints; however, readily available FPGA designs of DSP techniques operating at such wide bandwidths and with a large probe tone count are rare in the literature.

Common frequency synthesis schemes such as direct analog, direct digital synthesizer, and phase-locked loops are well-suited for generating single stable reference frequencies but do not address the challenges of synthesizing wide-band, multi-frequency signals\cite{lit_rev_Jeon2024, lit_rev_wu2013millimeter, lit_rev_Drechsel2019}. Over the past 20 years, polyphase filter bank (PFB) has been extensively used to analyze broadband signals and is followed by polyphase synthesis filter bank (PSB), or inverse PFB, to reconstruct channelized signals. Despite PFB's broad applications in radar, beam-forming, radio, and spectral analysis, there is little literature on utilizing PSB as the first component in a communication system to synthesize pre-defined spectra that are later modulated and analyzed. In 2020, Smith et al.\cite{lit_rev_OSPFBonZCU111_Smith} demonstrated an over-sampled PFB implemented on the same RFSoC used in this paper, achieving ultra-wide bandwidth channelization for MKID-based readouts, yet the polyphase synthesis aspect remained unexplored. Arnaldi's recent 2023 work\cite{lit_rev_CresFactorOpti_Arnaldi_2023} addressed crest factor optimization in multi-tone signal synthesis for low-temperature sensor readouts using the Newman algorithm. This optimization is crucial for maintaining the dynamic range of a DAC. However, the study lacks methods for synthesizing such signals, which is the focus of this study. The specific operational requirements of MKIDs and sub-mm telescopes create a niche need for the design and FPGA implementation of the OC-PSB.

The efficiency and flexibility of the PSB allow for an FPGA-friendly synthesizer design that adapts to the increasing demand for detector count and bandwidth. The results indicate that the introduction of the PSB not only lifts the readout bandwidth limitation but also enhances signal synthesis performance and enables next-generation readout capabilities through real-time signal editing within the FPGA fabric. This proof of concept opens up new MKID readout possibilities that can potentially advance sub-mm observations and enrich our understanding of the cosmos.

\section{Scope and Assumptions}
\label{sec: scope}
The synthesizer design in this paper is tailored to the 850-GHz instrument module on the Prime-Cam instrument. Still, it is adaptable to all MKID-based readouts and any wide-band multi-frequency synthesis applications. As the Prime-Cam instrument 850-GHz detector arrays are still under rapid development at the time of this study, the design specifications required by this study are based on results from preliminary studies and assumptions. For example, the SNR requirement derived in \autoref{sec: requirements} depends on the optimal biasing power required by MKIDs and this value can differ depending on the material choice and fabrication process of MKIDs. Other factors such as atmosphere loading can affect the biasing of MKIDs even during observation. In this regard, perhaps the flexibility and scalability of the synthesizer design are the most valuable contributions of this study.

Open-source Python libraries are utilized for algorithm design and simulations. MATLAB and Simulink, licensed by the University of British Columbia, are employed for the design and verification of FPGA firmware. The firmware building blocks in Simulink are based on libraries from Vitis Model Composer 2021.1 (Xilinx HDL toolbox), which require a license obtained through the purchase of the ZCU111 development kit.

\section{Synthesizer Performance Requirements and Considerations}
\label{sec: requirements}

The requirements are mainly influenced by the desired readout capabilities of the 850-GHz module, regarding operational functionality, measurement sensitivity, array limitations, and RFSoC capabilities. Due to the preliminary nature of the study, the requirements ensuring the MKIDs' theory of operation is strict while the efficiency and cost of implementation are desired but relaxed. A summary of requirements is listed in \cref{tab: requirements}.

\begin{table}[h!]
\centering
\begin{tabular}{|c|c|c|c|}
\hline
\multicolumn{4}{|c|}{Desired Performance Requirements} \\
\hline
Number of tones: 2048 & Bandwidth: 1.024GHz & SNR: 100dB & Resolution: 500 Hz\\
\hline
\multicolumn{4}{|c|}{Desired Functional Requirements} \\
\hline
\multicolumn{4}{|c|}{Scalability, real-time frequency/magnitude/phase adjustments} \\
\hline
\end{tabular}
\caption{Synthesizer requirements}
\label{tab: requirements}
\end{table}

As explained in \autoref{sec: intro}, the 850-GHz instrument module is pursuing reading out ~1200 MKIDs across 1.024 GHz of BW from a single coaxial cable. In terms of typical power-of-two hardware scaling, this translates to the requirement of 2048 tones. As we will see later, the 2048-point IFFT is the main component of the synthesizer and is directly related to the number of tones produced. A redundancy in probe tone count also ensures the utilization rate of MKIDs and accounts for the increasing multiplexing factor.

\begin{figure}[h]
    \centering
    \includegraphics[width=0.618\textwidth]{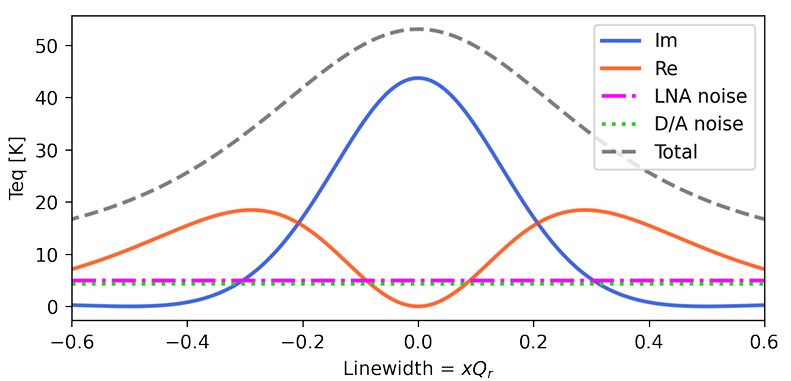}
    \caption{The effective noise temperature of an MKID in both quadratures, LNA and D/A noise, plotted against the shift from resonance in multiples of a linewidth. The dashed grey line represents the total noise summing D/A, LNA, imaginary, and real. Parameters used are derived from real-world measured in-flight median values for the BLAST-TNG experiment 850-GHz array and can be found in Sinclair et al. 2022 \cite{sinclair2022ccatprime}}
    \label{fig: noise stack}
\end{figure}

Prime-Cam instrument has the goal of detector noise limited readout performance. The preliminary study on the readout system noise stack\cite{sinclair2022ccatprime} is shown in \cref{fig: noise stack}.

The equivalent noise temperature method \cite{Sipola} is used to compare the noise contributions from key readout components, plotted against the shift
from resonance in multiples of a linewidth. MKIDs have complex transmissions and the ``Im" and ``Re" correspond to the equivalent output noise temperature in each quadrature. In a typical operation, the complex output signal is rotated such that only the "Im" component carries the science information, therefore, while the noise from the real part contributes to the total noise, it is neglected in the consideration for probe tone SNR. 

As shown in \cref{fig: noise stack}, without the noise contribution from the probe tone, the noise floor is largely dependent on the LNA and DAC noise which are around 5K. The readout noise is considered detector-limited if the detector output noise temperature is higher than other components. The probe tone noise temperature is desired to be around 5K, comparable to the LNA and DAC.

\Cref{eq: Biasing noise power calculation} converts the noise temperature of 5K to its measure in dBm/Hz,

\begin{equation}
    P_{noise} = 10 \cdot log_{10}(k \cdot T) = 10 \cdot log_{10}(\frac{1.38 \cdot 10^{-23} \cdot 5}{10^{-3}}) = -191.6115~[\textrm{dBm/Hz}]
    \label{eq: Biasing noise power calculation}
\end{equation}

where k is the Boltzmann constant, and T = 5 [K] is the desired noise temperature.

To operate MKIDs with maximum sensitivity, the bias tone power must be high enough without driving MKIDs to their bifurcation state \cite{Mauskopf2018}. The optimal tone power depends on the material choices and fabrication of MKIDs. A typical optimal bias tone power from previous telescopes is about -90 dBm/Hz. This gives us the required SNR in dB for the synthesizer probe tone, as shown in \cref{eq: Biasing SNR calculation}.

\begin{equation}
    SNR = -90~\textrm{dBm/Hz} - (-191.6115~\textrm{dBm/Hz}) \approx 100~\textrm{dB}
    \label{eq: Biasing SNR calculation}
\end{equation}

\Cref{fig: noise stack} also indicates the frequency resolution required by the synthesizer. To ensure a detector noise-dominated performance and reasonable detector sensitivity, the probe tone must land within -0.3 to 0.3 fractional line width around the resonance frequency which is roughly 500 Hz. With a frequency resolution of 500 Hz, without knowing exactly the detector frequency placement pre-fabrication, and in the worst case scenario, we can place a tone at a fractional line width that gives at least 20 K detector noise temperature.

The above requirements ensure that the MKIDs can operate at a region that maximizes their sensitivity and maintains detector-limited readout noise. Functional requirements beyond the theory of operation allow an adaptable design to the fast-evolving detector technology and promote enhanced readout capabilities.

In practice, the biasing of MKIDs is not as simple as placing a fixed probe tone at a fixed resonant frequency, or at least, operating in this mode limits the potential of MKID-based instrument measurements. To accommodate the challenge of shifting post-fabrication detector resonant frequency due to biasing and atmospheric loading, it is desired that the synthesizer can edit its frequency comb quickly during measurements to ``track" the shifting detectors, making sure that the readout system is always calibrated to the maximum sensitivity available.

Last but not least, With the current trend of increasing detector counts, the synthesizing algorithm must be scalable in bandwidth and number of tones in the sum of frequencies.

\section{Overlap-Channel Polyphase Synthesis Filter Bank (OC-PSB)}
\label{sec: PSB-design}
The proposed frequency comb synthesizer for the biasing of MKIDs consists of baseband DDS (CORDIC in vector rotation mode) and the polyphase synthesis filter bank (PSB) that transforms channel subband TDM signals into one wide-band FDM signal. The probe tone spectral shapes are defined at the baseband with a highly reduced sampling rate.

The derivation of a critically sampled (or maximally decimated) M-path polyphase filter bank (PFB) structure is well studied and documented as early as 1983 in the book "Multirate Digital Signal Processing" by Crochiere and Rabiner\cite{Multirate_Crochiere}, and most recently in the book "Multirate Signal Processing for Communication Systems 2ed" by Fredric Harris\cite{Multirate_Fred} in 2021. The two textbooks offer similar derivations but different viewpoints on the PFB. The polyphase synthesis filter bank is the inverse operation (or dual graph) of the polyphase filter bank. The PFB and PSB consist of three subsystems: commutator resampling, inverse fast Fourier transform (IFFT), and M-path polyphase FIR filtering.

The overlap-channel PSB (OC-PSB) is an extension of the critically sampled PSB by having double the input channels and IFFT length, with minor adjustments to the conventional sequence of operations. The design parameters of the implemented OC-PSB are summarized in \cref{tab: OVPSB}.
\begin{table}[h!]
\centering
\begin{tabular}{|l|l|}
\hline
\textbf{Parameter} & \textbf{Value} \\
\hline
Number of polyphase paths (M) & 1024 \\
\hline
Resampling factor (channel to output) & 1 to M \\
\hline
Number of channels & 2048 \\
\hline
Number of prototype filter taps & $16 \times 1024$ \\
\hline
Prototype filter method & Windowed sinc \\
\hline
Prototype filter window function & Blackman–Harris \\
\hline
\end{tabular}
\caption{Design parameters for OC-PSB}
\label{tab: OVPSB}
\end{table}

\begin{figure}[h]
    \centering
    \includegraphics[width=0.618\textwidth]{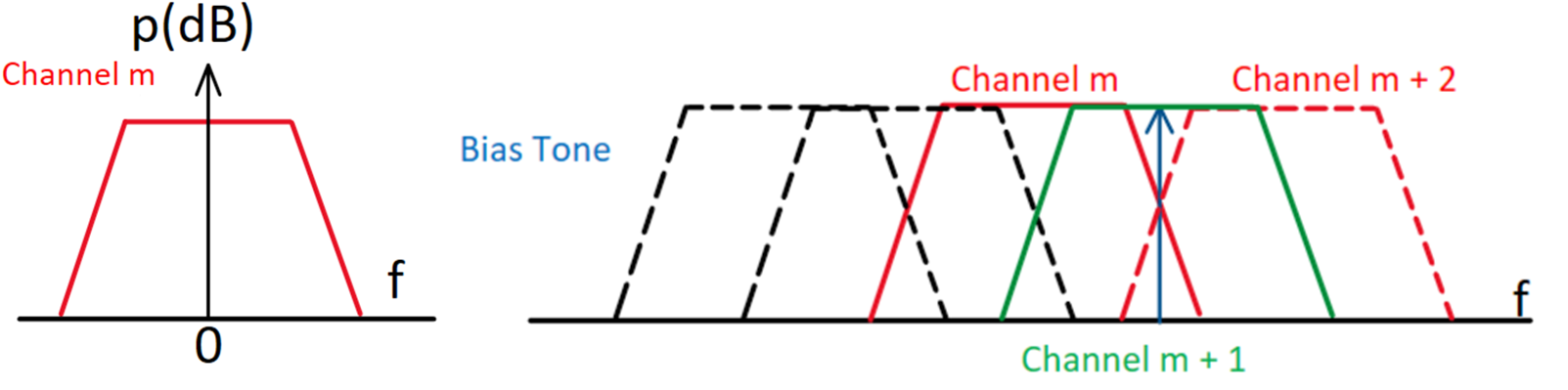}
    \caption{OC-PSB Channel frequency response and bias tone placement}
    \label{fig: OVchannelDrawing}
\end{figure}

\Cref{fig: OVchannelDrawing} illustrates the channel frequency response of the OC-PSB. The wide band with sampling rate $f_s$ is equally divided into K channels with even indexing, illustrated in red. These even-indexed channels constitute a critically sampled PSB. The same PSB with channel centers shifted by $\frac{f_s}{2K}$ become odd-indexed channels overlapping on the even-indexed channels, shown in green. The benefit of overlapped channels becomes apparent when a probe tone (shown in blue) located near the edge or transition of even indexed channels is synthesized. Unless the prototype FIR filter of the PSB has a length approaching infinity, the channel band-pass transition cannot be narrow enough to account for a detector resonant frequency landing near the channel edge. But with the additional channel m+1, the same frequency can be synthesized with high SNR at a channel center instead. Separately, the even and odd indexed channels are known as ``channelizers with even and odd indexed bin centers"\cite{Multirate_Fred}.

\subsection{OC-PSB: Sequence of Operation}
\begin{figure}[h!]
    \centering
    \begin{minipage}{0.45\textwidth}
        \centering
        \includegraphics[width=\textwidth]{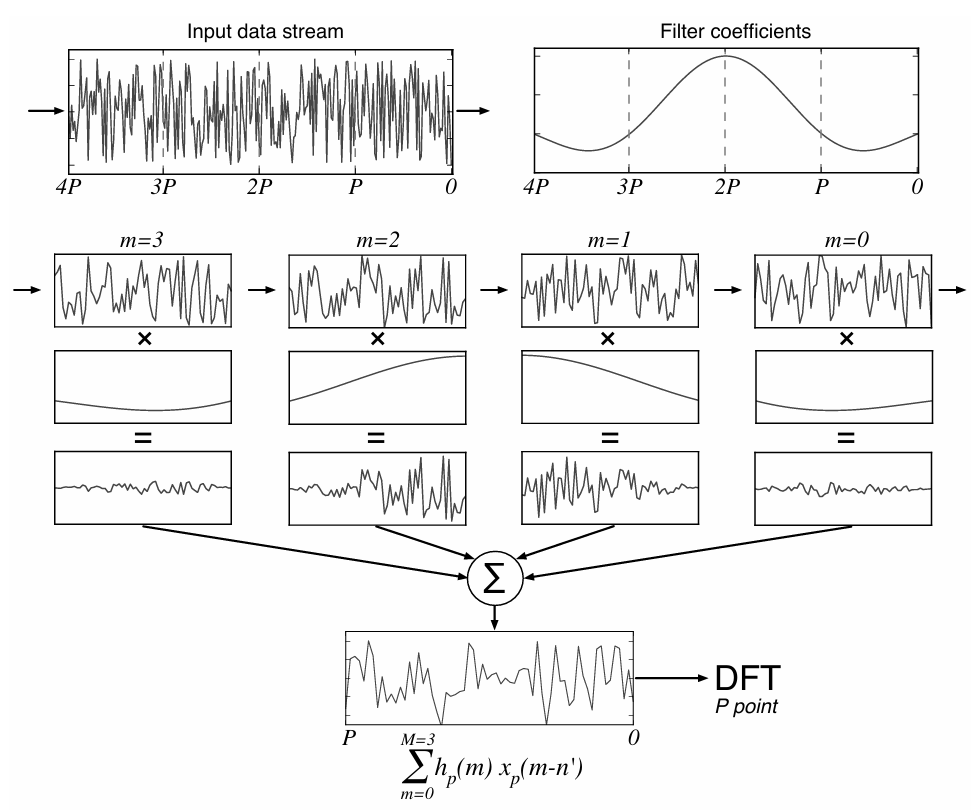}
        \caption{Graphical representation of a critically sampled PFB by Danny Price\cite{Danny_PFB_in_astronomy}, with 64 channels (P) and 4 taps per channel (M). Data are read in blocks of P until M x P samples are buffered. The data and filter coefficients are then split into 4 taps, multiplied, and summed. After summation, a P-point DFT is computed for outputs and another P input samples are loaded.}
        \label{fig: PFBillustration}
    \end{minipage}
    \hfill
    \begin{minipage}{0.45\textwidth}
        \centering
        \includegraphics[width=\textwidth]{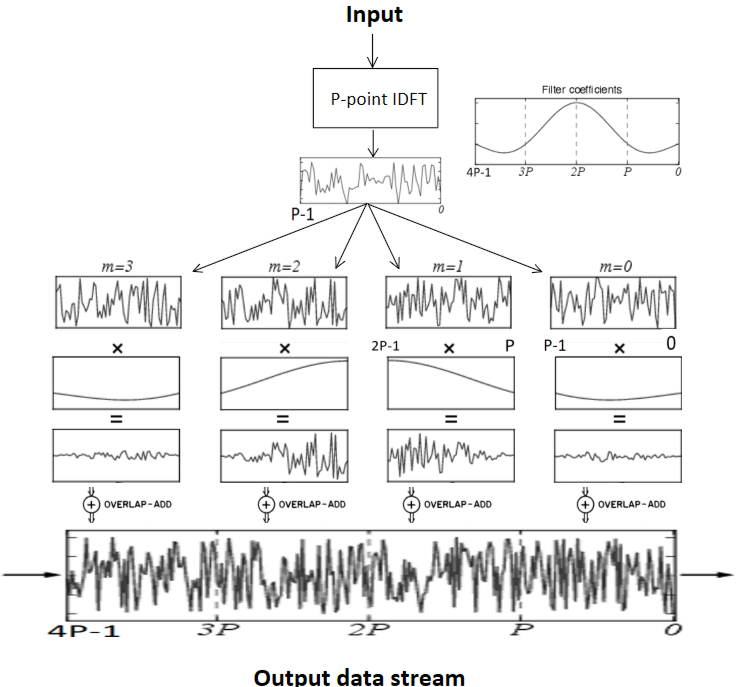}
        \caption{Graphical representation of a critically sampled PSB, derived as the dual graph of Danny Price's PFB illustration. A block of P input goes through a P-point IFFT before periodically extended by a factor of M, the M x P samples then multiply with the re-petitioned filter coefficients. The blocks then enter an accumulator buffer of length M x P, summing with the previous buffer contents. Lastly, the buffer shifts by P, the block shifting out of the buffer are outputs, and zeros are filled in the buffer from the left. Another block of P will be processed at input, and the shifted content of the buffer becomes the previous buffer contents.}
        \label{fig: CSPSBflow}
    \end{minipage}
\end{figure}

\begin{figure}[h!]
  \begin{minipage}[b]{0.45\linewidth}
    \centering
    \includegraphics[width=\linewidth]{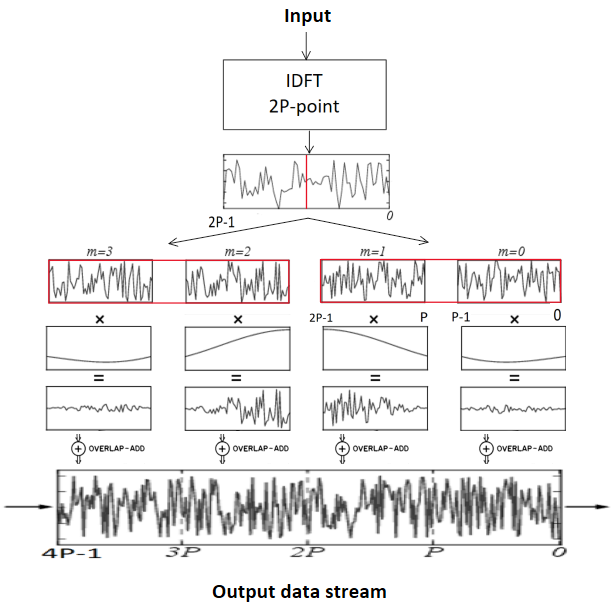}
    \caption{Graphical representation of the OC-PSB, extended from the critically sampled sequence of operations. The input becomes a block of 2P, going through a 2P-point IDFT, the input vectors at the odd-indexed bins are rotated by 180 degrees. The block of 2P is then periodically extended by a factor of M/2 to obtain the M x P sample blocks. The sequence of operation after this will be identical to that of a critically sampled case}
    \label{fig: OVPSBflow}
  \end{minipage}
  \hspace{0.05\linewidth}
  \begin{minipage}[b]{0.45\linewidth}
    \centering
    \includegraphics[width=\linewidth]{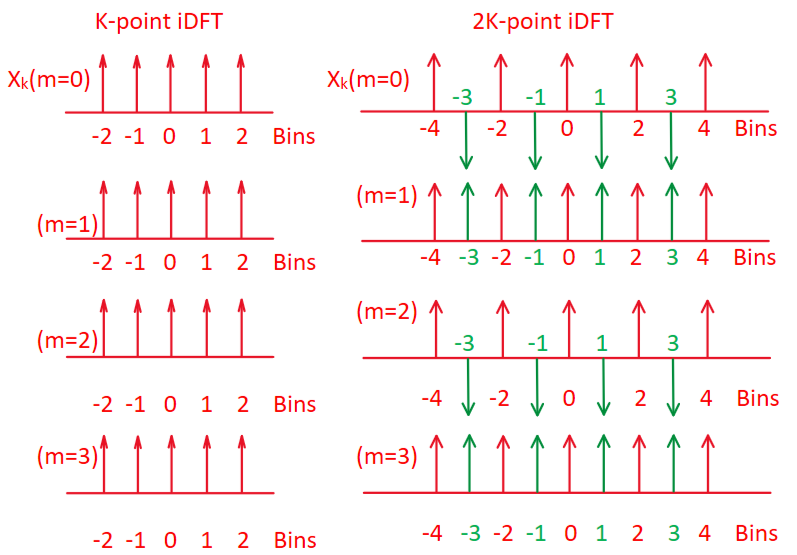}
    \vspace{20pt}
    \caption{Alternating signs at the odd-indexed IDFT input bins. This sign inverting, or 180-degree phase shift corresponds to the shift-by-half the channel spacing channelizer with odd-indexed bins}
    \label{fig: alternating input}
    \vspace{55pt}
  \end{minipage}
\end{figure}

The OC-PSB sequence of operations is derived from the ``graphical representation of a polyphase filterbank" by Danny Price, shown in \cref{fig: PFBillustration}. This graph is a critically sampled PFB and is widely used by the Collaboration for Astronomy Signal Processing and Electronics Research (CASPER). The critically sampled PSB is derived as the dual graph of PFB, shown in \cref{fig: CSPSBflow}, and the overlapping channels are introduced by the modifications explained in \cref{fig: OVPSBflow} and \cref{fig: alternating input}.

\subsection{OC-PSB Performance}

To highlight the overlapping channel design, the performance of the OC-PSB is demonstrated by a comparison between a single probe tone synthesized by a critically sampled design at a channel edge, and by the OC-PSB at the adjacent odd-indexed channel center. Since the channel response is consistent across all channels of a PSB, the performance of the single tone is representative of any other tone. The critically sampled design has half the number of channels and a filter length of 64 x 1024—four times that of the overlapping design—to minimize the channel band-pass transition band. Python is used to showcase the theoretical performance of the PSB design through floating-point simulations.

\begin{figure}[h]
  \begin{minipage}[b]{0.45\linewidth}
    \centering
    \includegraphics[width=\linewidth]{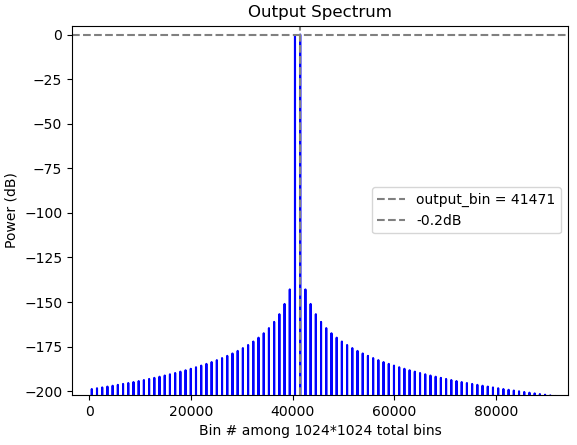}
    \caption{Output spectrum of a probe tone synthesized at a channel edge by a critically sampled PSB}
    \label{fig: CSedgeOPspectrum}
  \end{minipage}
  \hspace{0.05\linewidth}
  \begin{minipage}[b]{0.48\linewidth}
    \centering
    \includegraphics[width=\linewidth]{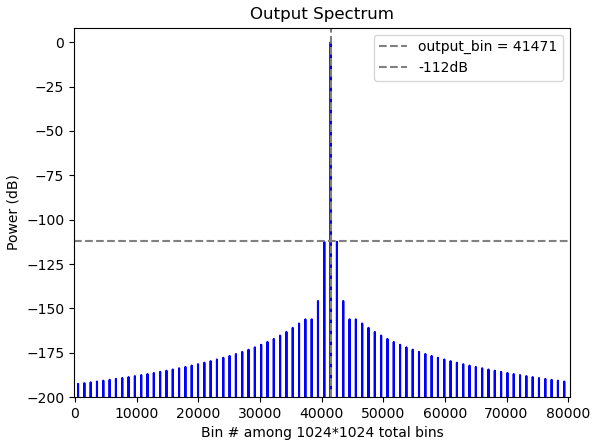}
    \caption{Output spectrum of the same probe tone synthesized at the adjacent odd-indexed channel center by OC-PSB}
    \label{fig: OVoddCenterOPspec}
  \end{minipage}
\end{figure}

We can see that the worst image power is almost identical to the target probe tone in the critically sampled design (\cref{fig: CSedgeOPspectrum}), whereas the images are well below -100 dB if the same tone is synthesized in the additional overlapping channel by the OC-PSB(\cref{fig: OVoddCenterOPspec}).

The output spectrum is determined by the channel frequency response, which is determined by the prototype FIR filter.

\begin{figure}[h!]
  \begin{minipage}[b]{0.45\linewidth}
    \centering
    \includegraphics[width=\linewidth]{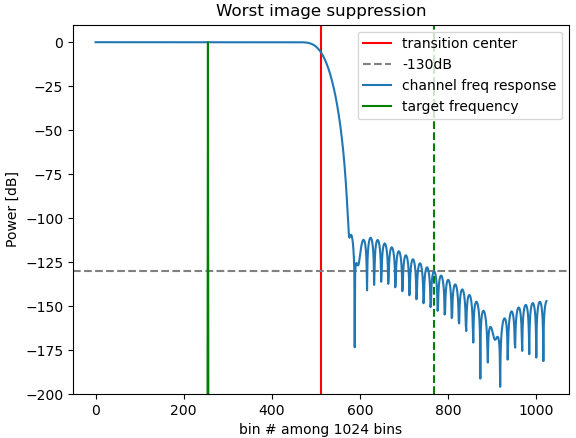}
    \caption{Blue: Channel shape of the critically sampled PSB with prototype filter length = $64 \times 1024$. Green: probe tone at input\_bin = 256. Red: channel edge/transition center}
    \label{fig: CSmidIPspectrum}
  \end{minipage}
  \hspace{0.05\linewidth}
  \begin{minipage}[b]{0.45\linewidth}
    \centering
    \includegraphics[width=\linewidth]{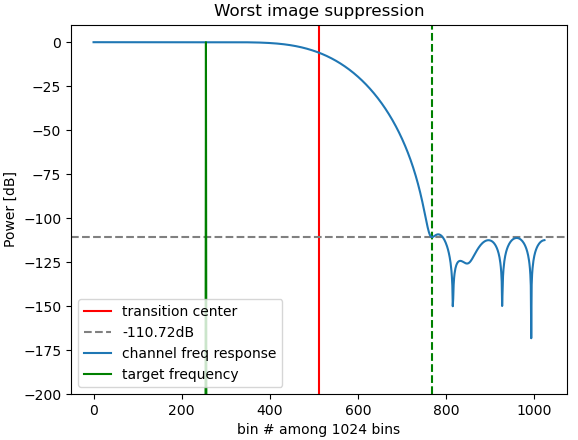}
    \caption{Blue: Channel shape of the OC-PSB with prototype filter length = $16 \times 1024$. Green: probe tone at input\_bin = 256. Red: channel edge/transition center}
    \label{fig: OVchannelFR}
  \end{minipage}
\end{figure}

The intricacy of the PSB design is perhaps condensed in \cref{fig: CSmidIPspectrum} and \cref{fig: OVchannelFR}. They reveal the channel frequency response of the critically sampled case and OC-PSB obtained by a 1024-point FFT, along with the input spectrum in the channel, and the worst spectral image location and power.
The blue curve is the right half of a channel frequency response at the baseband centered at DC. If expanded by symmetry, it is clear that the channel passband occupies 1024 bins from bin -512 to 511, that is 1/1024 of the output bandwidth defined by the $1024 \times 1024$ bins in \cref{fig: CSedgeOPspectrum} and \cref{fig: OVoddCenterOPspec}. The shape of both channels is determined by the Blackman-Harris window function, resulting in a transition bandwidth of roughly 80 bins (if $f_s = 1.024 GHz$, the transition BW = 78125 Hz) for the critically sampled case, and much wider for the overlapped case. Stop band attenuations are higher than 100 dB. A narrower transition BW would require more number of taps in the prototype LPF. Blackman-Harris window function is chosen for its narrow transition band and high stop-band attenuation.
The solid red vertical line is the middle of two channels or the channel edge. If the blue curve is channel k, then the frequency response of channel k+2 would be the blue curve mirrored with respect to the red line. There are a total of 1024 channels with the same shape in the critically sampled PSB design side by side with no overlapping pass band. And for the overlapped design another set of 1024 channels is present as explained with \cref{fig: OVchannelDrawing}.
The green Dirac-delta is the input spectrum of channel k at the baseband. In this case, it is a tone located in between the channel center (at 0) and the edge (at the red line) with input\_bin = 256 (in terms of a 1024-point FFT within channel sub-band).
Before the channel band-pass filtering, there are 1024 equal-spaced replicas or images of the green tone. Since this tone 
is on the right side of channel k, the image closest to channel k is located in channel k-1 on the left side of the channel, suffering the worst attenuation by the stopband of channel k. Although this image is located at bin -769, by symmetry, the cross-over of the closest image and channel k stopband is plotted as the cross-over of the blue curve and mirror image of the green tone with respect to the red line. From this we can see that when the target frequency approaches the edge of the channel, the closest image approaches the transition band of the channel, leading to unsatisfying noise power.

One notable difference is that the transition band is much wider for the OC-PSB due to $1/4$ reduced filter taps, however, this is allowed in the OC-PSB because any tone located beyond bin 255 or below bin -256 can be instead synthesized in the adjacent channel within bin -256 to 255, yielding image suppression higher than 100dB. The OC-PSB prototype filter design fills the redundant bandwidth with the transition band and, therefore largely relaxes the requirement on filter length, reducing hardware utilization.

\subsection{Design Limitations}
Although the OC-PSB can synthesize 2048 tones simultaneously with incredible frequency resolution (shown in \autoref{sec: firmware}), it currently only supports one tone per channel. More than one tone per channel is not only more challenging to simulate but as shown later, requires more sophisticated baseband waveform synthesis in hardware. Additionally, two tones from adjacent channels must be at least one channel spacing apart to ensure that no tones land in the transition bandwidth as shown in \cref{fig: OVchannelFR}. This is a trade-off between reduced filter taps and minimum tone spacing. If tone spacing smaller than one channel spacing is required, then the OC-PSB must have an increased prototype filter length.

\subsection{Conclusion}
The Python simulation performances indicate that the proposed design of an overlapped-channel PSB will meet all of the requirements listed in \autoref{sec: requirements}, with minor limitations that may or may not affect the MKID operation. The bandwidth of the OC-PSB or the maximum sampling frequency is limited by the FPGA implementation. With this, the study moves to the FPGA firmware design of the above-mentioned OC-PSB algorithm.

\section{FPGA Firmware Implementation}
\label{sec: firmware}
The synthesizer design is implemented in the Zynq UltraScale+ RFSoC ZCU111 Evaluation Kit. It utilizes the Xilinx HDL toolbox in Vitis Model Composer 2020.1, integrated with the Simulink environment. The toolbox is a library of firmware building blocks that can be simulated in Simulink and later converted to synthesizable hardware description language (HDL) for the FPGA implementation in the Vivado design suite. A combination of Python, Matlab/Simulink, and Vivado design suite is used for the simulation and verification.

The synthesized waveform is in 16-bit fixed point format. From typical assumptions for a DAC, the best achievable SNR can be estimated from the number of bits N of the DAC:
\begin{equation}
    SNR \approx 6.02N + 1.76 ~\textrm{dB} = 6.02 \cdot 16 + 1.76 = 98.08 ~\textrm{dB}
    \label{eq: SNRestimate}
\end{equation}

\begin{figure}[h]
  \begin{minipage}[b]{0.25\linewidth}
    \centering
    \includegraphics[width=\linewidth]{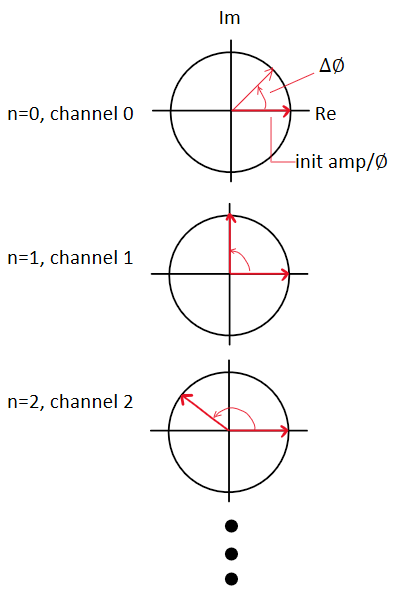}
    \caption{CORDIC illustration with unit circle}
    \label{fig: VectorRotation}
  \end{minipage}
  \hspace{0.05\linewidth}
  \begin{minipage}[b]{0.6\linewidth}
    \centering
    \includegraphics[width=\linewidth]{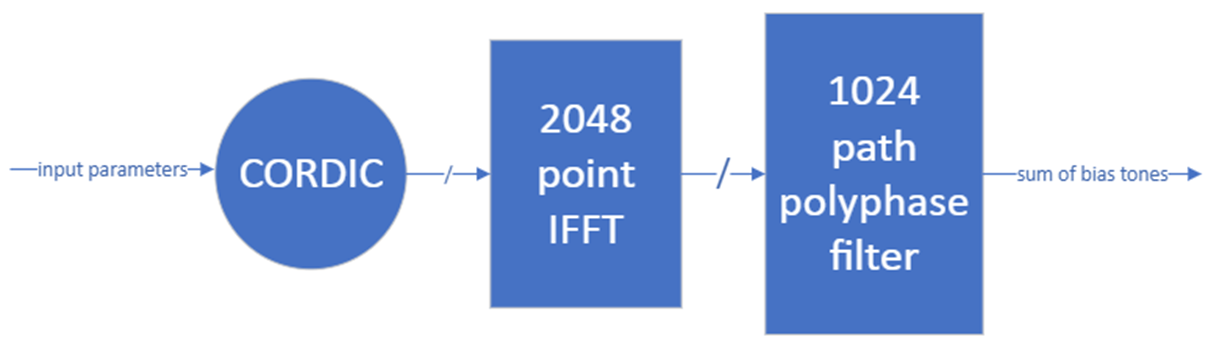}
    \vspace{10pt}
    \caption{High-level view of OC-PSB firmware design}
    \label{fig: HWHLblockDiagram}
  \end{minipage}
\end{figure}

The synthesizer is composed of two main modules, the CORDIC module that computes the input TDM waveforms for the channel inputs to the OC-PSB, and the OC-PSB module which consists of an IFFT and M-path polyphase filtering stage, as shown in \cref{fig: HWHLblockDiagram}.

One straightforward way of computing a digital complex waveform is to employ CORDIC in its vector rotation mode, with a constant $\Delta \phi$ to control the frequency, and an initial vector to control the magnitude and phase as shown in \cref{fig: VectorRotation}.
Note that only three of the 2048 channel inputs are shown. The TDM nature of the input waveforms allows the reuse of one CORDIC operating at $f_s$ to compute the input to 1024 channels. The sampling rate for each channel is $f_s/1024$. A total of two CORDICs are used for 2048 channels.

The IFFT module used for the OC-PSB performs a standard streaming 2048-point IFFT.

Before the outputs from IFFT are supplied to the polyphase filtering stage, they must be buffered, re-ordered, and replicated by a reordering buffer implemented with B-RAM to account for the periodic extension operations described in \cref{fig: OVPSBflow}.

\begin{figure}[h]
    \centering
    \includegraphics[width=0.618\textwidth]{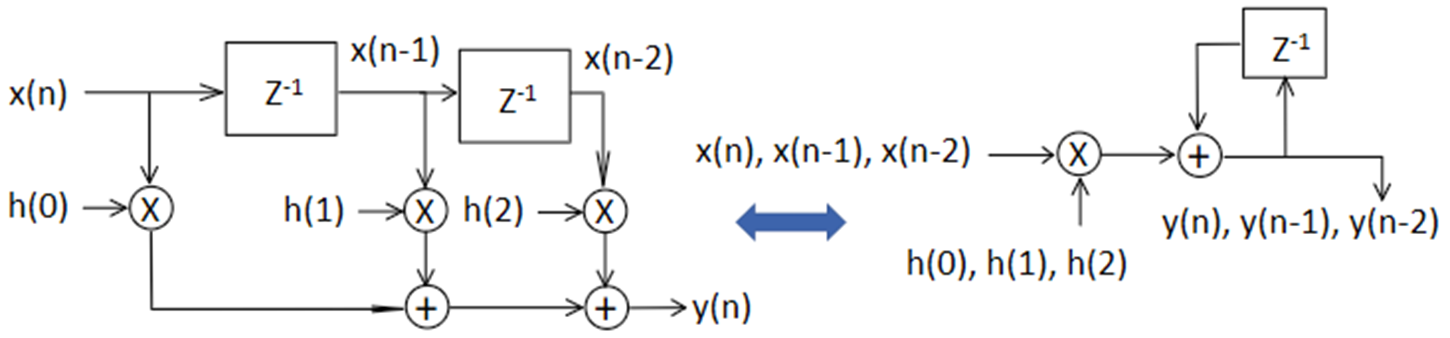}
    \caption{Standard 3-tap FIR structure(left), area-optimized (right)}
    \label{fig: FIRareaOptimized}
\end{figure}

\begin{figure}[h]
    \centering
    \includegraphics[width=0.618\textwidth]{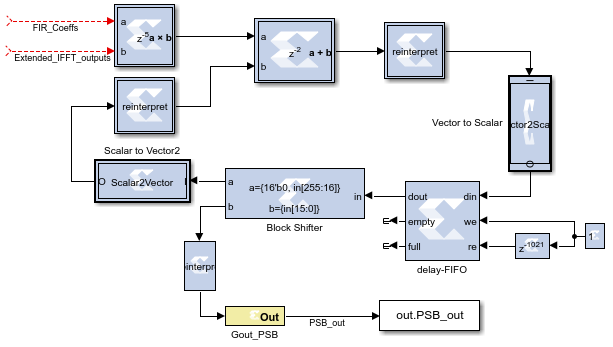}
    \caption{1024-path polyphase filtering stage implementation in Simulink}
    \label{fig: SimulinkFIR}
\end{figure}

The polyphase FIR filtering stage employs an area-optimized approach as illustrated in \cref{fig: FIRareaOptimized}.
In this configuration, each tap in each channel uses one multiplier and one adder for each of the real and imaginary components. The Simulink implementation of the 1024-path polyphase filtering is shown in \cref{fig: SimulinkFIR}.
The blocks with double-lined frames are operating in the SSR=16 mode, for instance, the adder shown in \cref{fig: SimulinkFIR} are 16 adders in parallel with vectorized inputs and outputs. The number of parallels is equal to the number of taps per channel. As illustrated in \cref{fig: OVPSBflow}, at cycle 0, the first samples of every polyphase path are delivered to the first taps in each re-petitioned sub-band filter. The 1024-path is operating in a TDM fashion.
A block shifter implemented with a bit brasher is used to shift the valid output samples out of the "overlap-add" shifting accumulator, implemented with a feedback loop.

\subsection{OC-PSB FPGA Implementation Performance}
\begin{figure}[h!]
    \centering
    \includegraphics[width=0.618\textwidth]{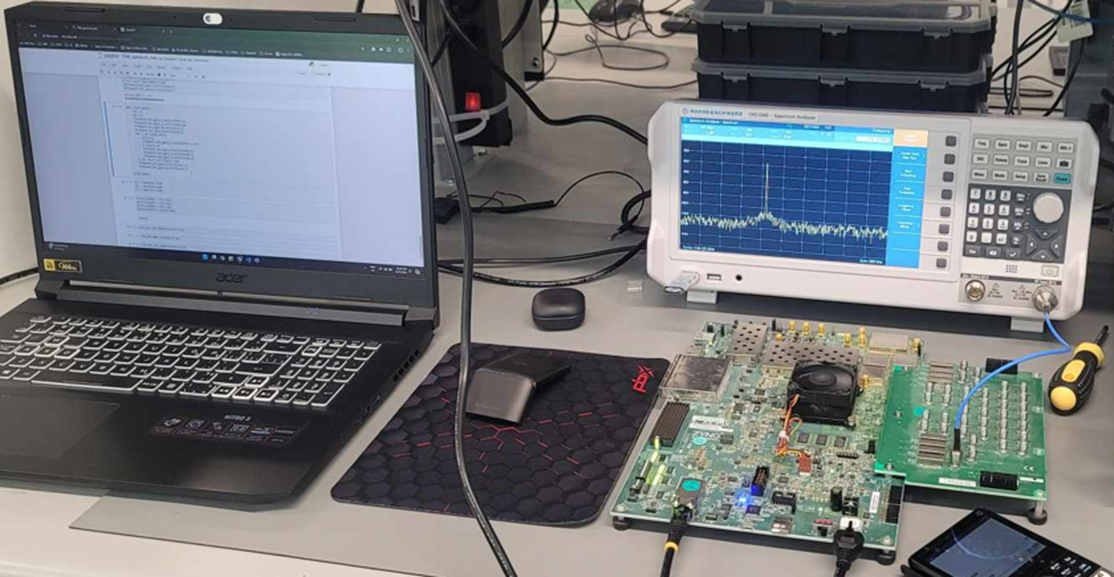}
    \caption{Result measurement set up, control computer (left), spectrum 
analyzer (top right), Xilinx ZCU111 development board (bottom right)}
    \label{fig: SpecMeasurementSetup}
\end{figure}

The final firmware design is verified in a combination of Python, Simulink, and Vivado simulations before it is implemented in the ZCU111 development board. High-speed DACs on board are used to convert digital output signals into measurable electrical signals by a Rohde and Schwarz FPC1500 Spectrum Analyzer. \Cref{fig: SpecMeasurementSetup} shows the measurement setup with control computers adjusting the frequency comb during synthesizer operation via an Ethernet cable connected to the RFSoC, and the RFSoC output pin connected to the spectrum analyzer.

\begin{figure}[h]
  \begin{minipage}[b]{0.45\linewidth}
    \centering
    \includegraphics[width=\linewidth]{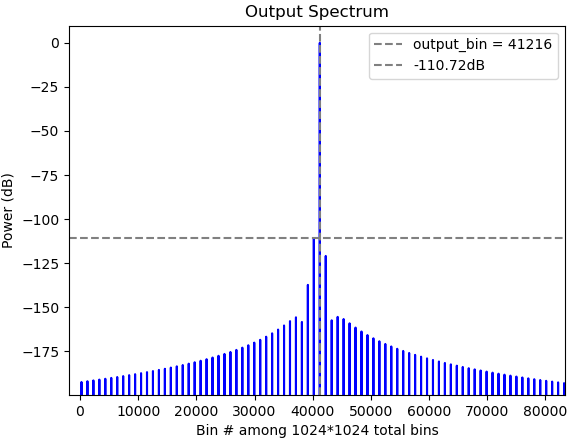}
  \end{minipage}
  \hspace{0.05\linewidth}
  \begin{minipage}[b]{0.49\linewidth}
    \centering
    \includegraphics[width=\linewidth]{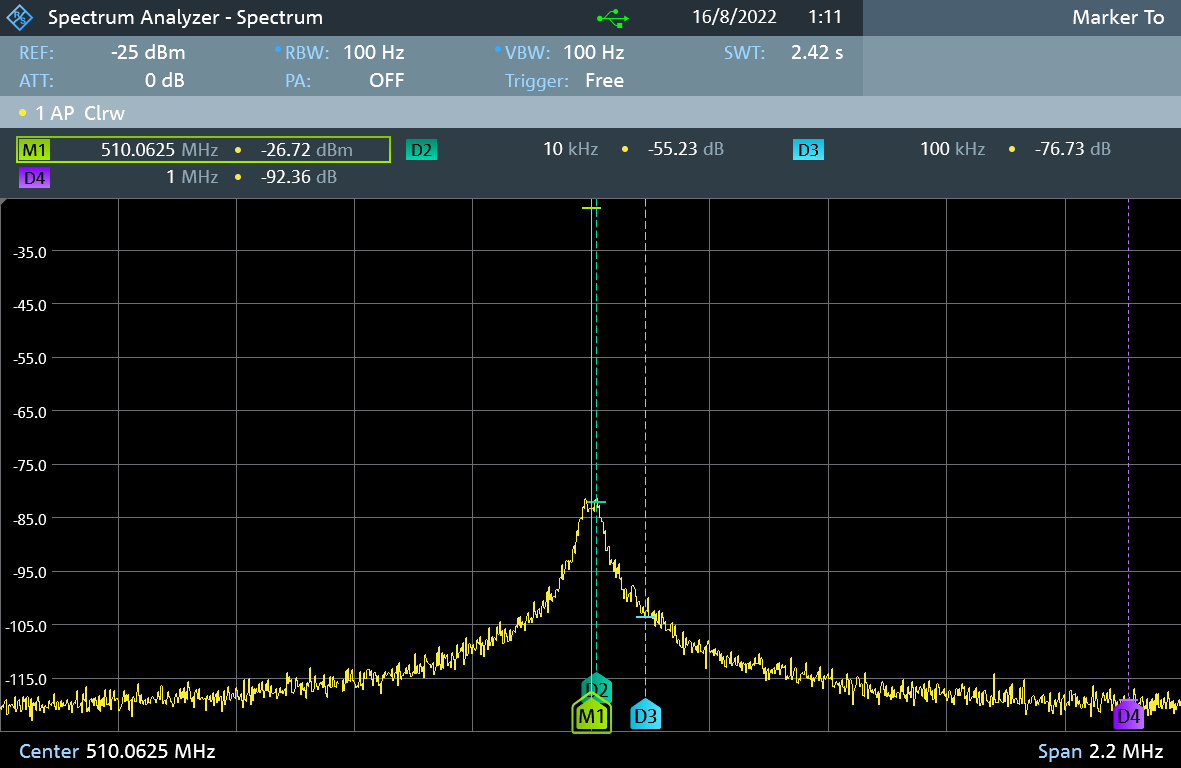}
    \hspace{15pt}
  \end{minipage}
  \caption{Python simulation (left) and spectrum analyzer measurement (right) for channel 80, input\_bin 256,  the same as the tone defined in \cref{fig: OVchannelFR}}
  \label{fig: SNRcomparison}
\end{figure}

The same tone defined in \autoref{sec: PSB-design} \cref{fig: OVchannelFR} is measured with the spectrum analyzer for comparison, showing in \cref{fig: SNRcomparison}.
Note that the output from the RFSoC is mixed with a numerically controlled local oscillator (NCLO) and therefore exhibits a frequency shift of 500 MHz.
The output frequency can be calculated using \cref{eq: measurement frequency calculation}:
\begin{equation}
    f_{out} = (1024 \cdot \frac{80}{2} + 256) \cdot \frac{256\cdot10^6}{1024^2} + 500\cdot10^6 = 510.0625~[\textrm{MHz}]
    \label{eq: measurement frequency calculation}
\end{equation}
Where 80 is the input channel index (among a total of 2048 channels) and 256 is the input\_bin defined in a 1024-point FFT. The FPGA has clock rate of 256 MHz and the output bins are defined in a $1024 \times 1024$-point FFT. The 500 MHz is the frequency shift by the NCLO. Note that bin 80 for a 2048-point IFFT is equivalent to bin 40 in a 1024-point IFFT.

Marker 1 is set at the peak of the spectrum, showing the frequency and absolute power of the target tone, matching 510.0625 MHz with a resolution bandwidth (RBW) of 100 Hz, while markers D2, D3, and D4 show the frequency and power offset from the target tone. The SNRs measured at 10k, 100k, and 1M Hz away from the peak are 55.23dB, 76.73dB, and 92.36dB respectively. The output spectrum has Lorentzian-like shaped wings around the target tone, with a noise floor nearing the maximum SNR estimated in \cref{eq: SNRestimate}. The theoretically estimated spectral images are low enough to be buried under the Lorentzian-like wings.

\begin{figure}[h]
  \begin{minipage}[b]{0.45\linewidth}
    \centering
    \includegraphics[width=\linewidth]{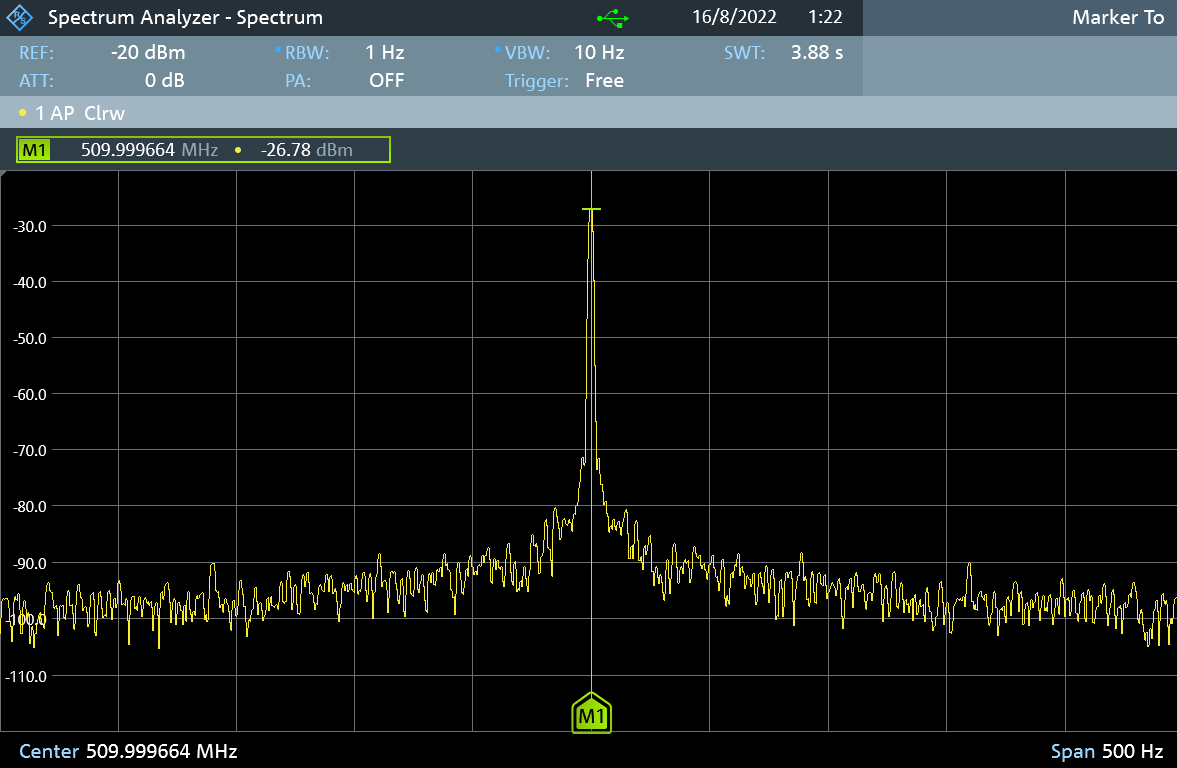}
  \end{minipage}
  \hspace{0.05\linewidth}
  \begin{minipage}[b]{0.45\linewidth}
    \centering
    \includegraphics[width=\linewidth]{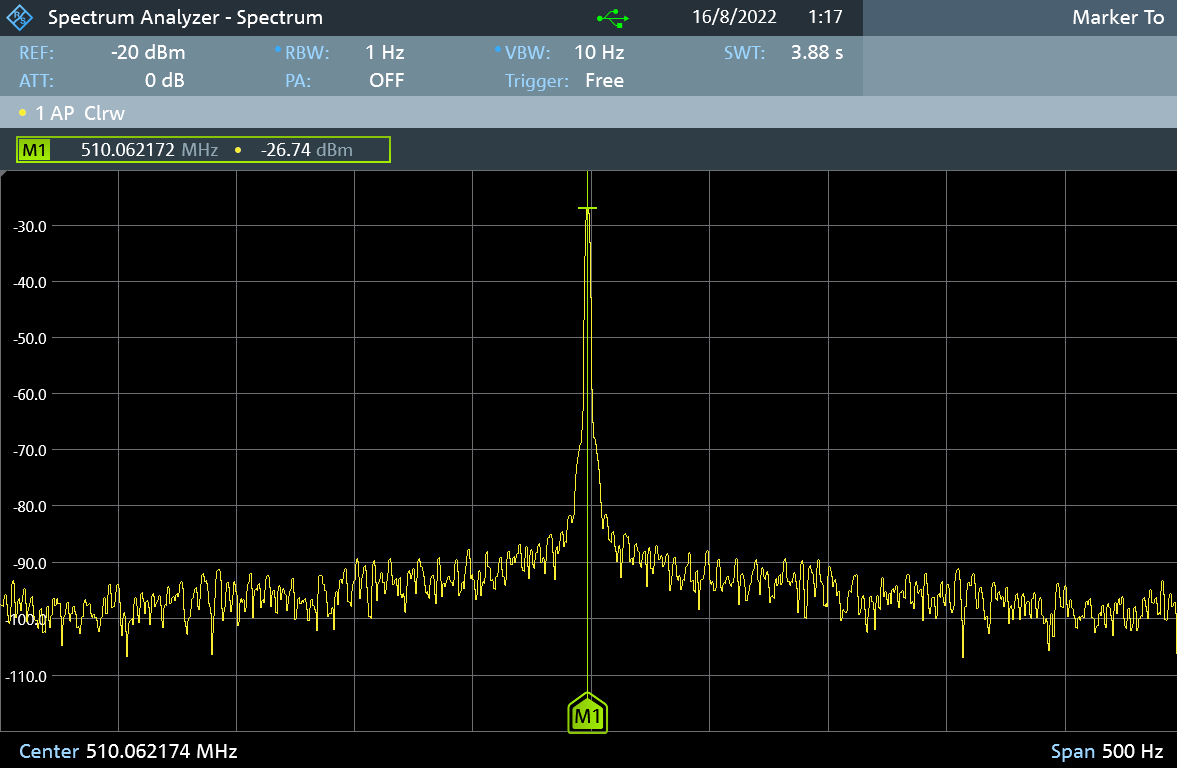}
  \end{minipage}
  \caption{Two tones with 256 output\_bin, or 62.5 kHz spacing measured at RWB = 1Hz}
  \label{fig: accuracy}
\end{figure}

\Cref{fig: accuracy} demonstrates the frequency accuracy of the synthesizer by zooming in on the previously measured tone and a tone in the same channel at the channel center.
The measured probe tone peak frequency does not match the target frequency exactly because there is a known frequency offset with the RFSoC clock frequency. This offset is likely temperature-related and is inherent in the physical clock of the RFSoC. However, it is a constant offset during operation that can be calibrated.
Nonetheless, the synthesizer accuracy (evaluated by measuring the frequency error, defined as the difference between the intended frequency and the measured output frequency) can be measured by the frequency difference of the two tones (256 output\_bins, or 62.5 kHz apart) measured in \cref{fig: accuracy}. Comparing the theoretical frequency difference and the measured, the frequency error is less than 10 Hz.

\begin{figure}[h!]
    \centering
    \includegraphics[width=0.5\textwidth]{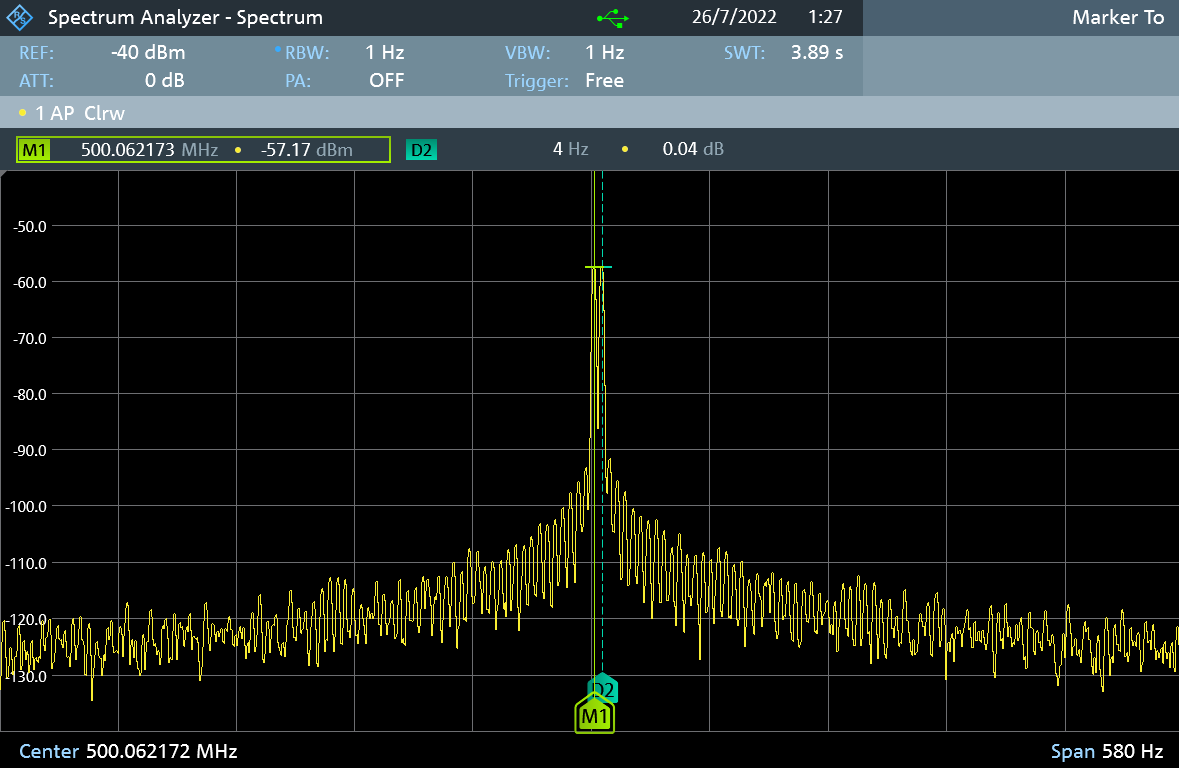}
    \caption{Measured frequency resolution of the synthesizer}
    \label{fig: MeasResolution}
\end{figure}

\Cref{fig: MeasResolution} illustrates the system's frequency resolution capability by displaying two tones that are the closest possible frequencies the system can resolve, demonstrating a minimum resolvable frequency separation of 4 Hz. The PSB itself does not have inherent frequency resolution, but in hardware, this resolution comes from the quantization in the CORDIC computing input waveforms. A fine frequency resolution can help place the bias tone very close to the center of the detector resonant frequency which brings the optimal sensitivity and dynamic range.

\begin{figure}[h!]
  \begin{minipage}[b]{0.45\linewidth}
    \centering
    \includegraphics[width=\linewidth]{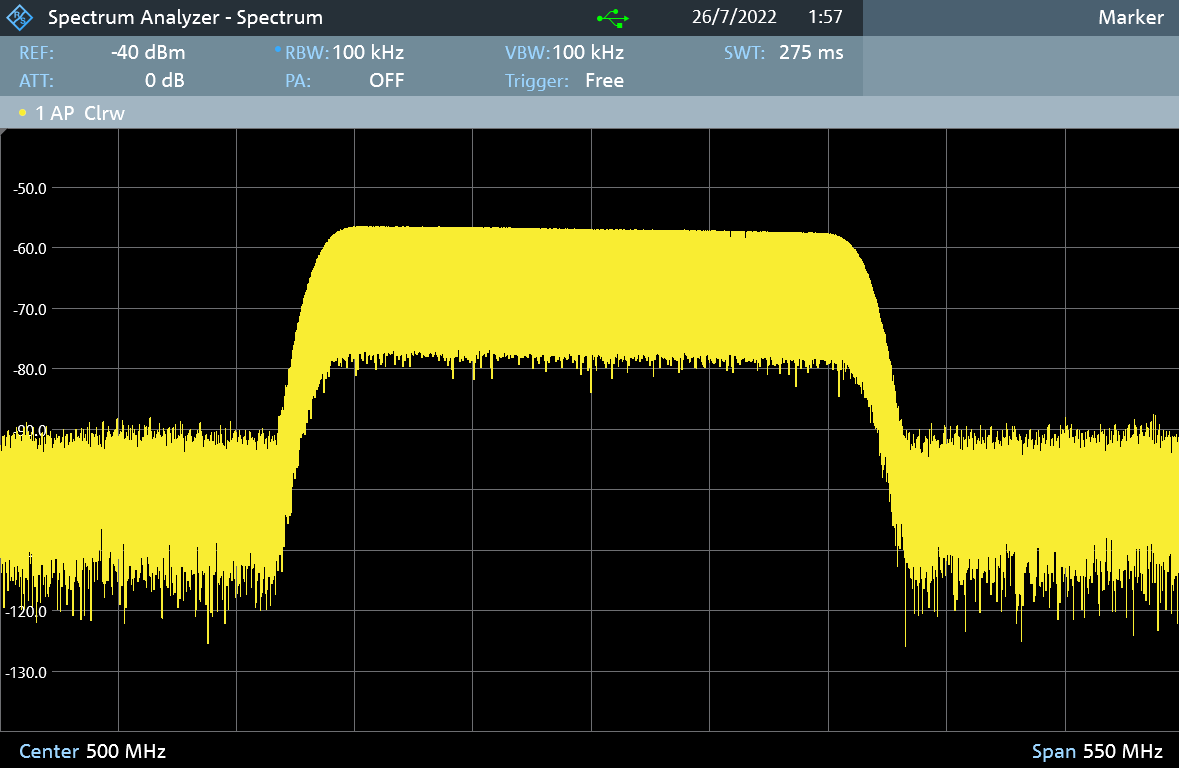}
  \end{minipage}
  \hspace{0.05\linewidth}
  \begin{minipage}[b]{0.45\linewidth}
    \centering
    \includegraphics[width=\linewidth]{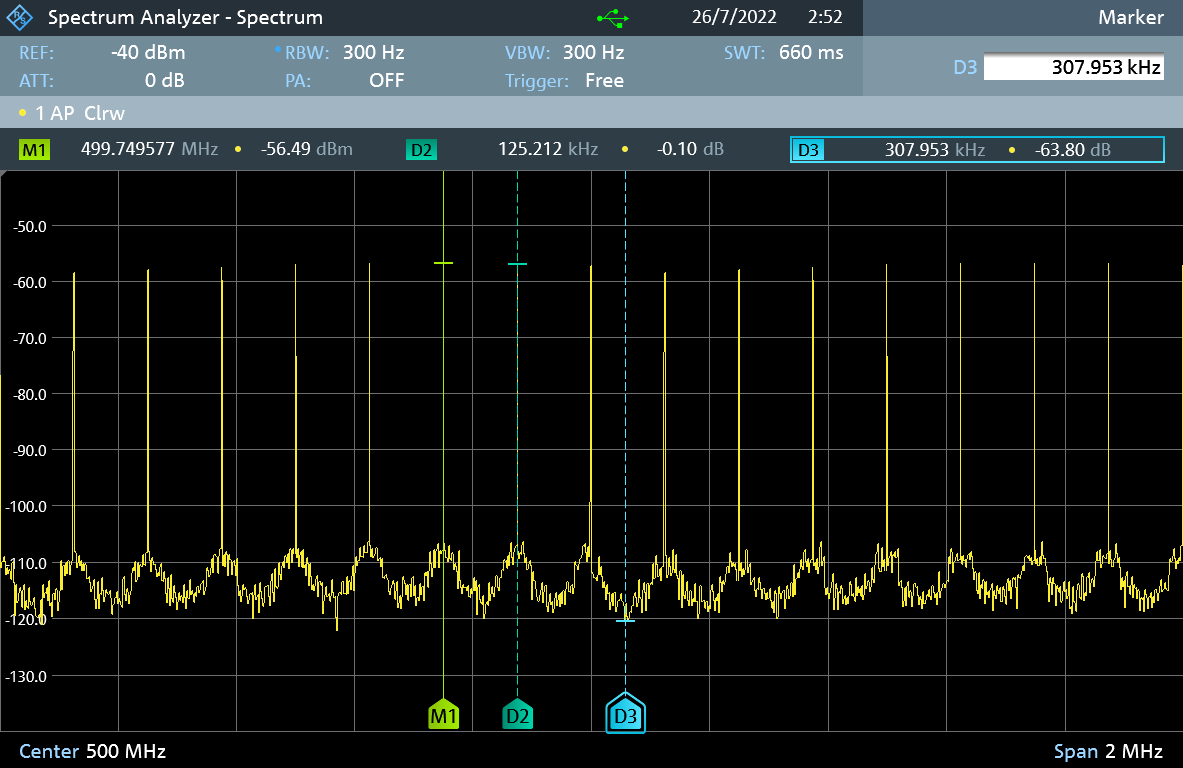}
  \end{minipage}
  \caption{All 2048 channels turned on with equal spaced tones, zoomed in on the right}
  \label{fig: allChannel}
\end{figure}

\Cref{fig: allChannel} demonstrates the synthesizer producing 2048 equal-spaced tones by turning on all the channels across the 256 MHz bandwidth.
The frequency spacing matches the channel spacing of $\frac{256 \textrm{MHz}}{2048} = 125 \textrm{kHz}$. The frequency comb has reduced equal magnitudes and randomized phase for each tone to prevent overflow by managing crest factor. The noise floor will be raised to the crossover of two closely spaced wings.

\newpage
The measured performance of the synthesizer implementation is summarized below:
\begin{itemize}
    \item Number of channels: 2048
    \item Number of tones per channel: 1
    \item Frequency error: $<$ 10 Hz
    \item Frequency resolution: 4 Hz
    \item SNR (Single tone) at 1 MHz: 92.36 dB
    \item Bandwidth: 256 MHz
    \begin{itemize}
        \item Constrained by maximum FPGA clock rate, but can be scaled to multiples of 256 MHz by increasing the parallel factor in fabric (eg. parallel by 4 to achieve $f_s = 1.024~GHz$)
    \end{itemize}
    \item Real-time ($< 20 \mu$s) frequency/magnitude/phase adjustment
\end{itemize}

\subsection{FPGA Resource Utilization}
\begin{figure}[h]
  \begin{minipage}[b]{0.45\linewidth}
    \centering
    \includegraphics[width=\linewidth]{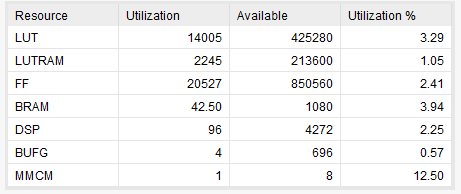}
    \vspace{10pt}
  \end{minipage}
  \hspace{0.05\linewidth}
  \begin{minipage}[b]{0.45\linewidth}
    \centering
    \includegraphics[width=\linewidth]{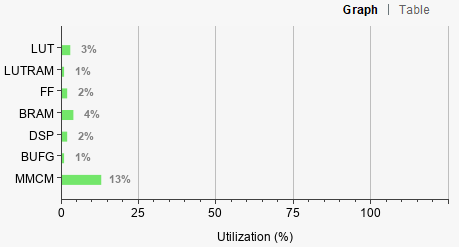}
  \end{minipage}
  \caption{Left: utilization table; right: utilization graph}
  \label{fig: UtilData}
\end{figure}

The post-implementation FPGA resource utilization of the synthesizer is reported by Vivado in \cref{fig: UtilData}.
The reports include the synthesizer module as well as the necessary interface to the ZYNQ processor and DACs aboard the Xilinx RFSoC. The high MMCM utilization can be ignored due to large number of testing ports existing in the synthesized test design. 

\begin{figure}[h]
    \centering
    \includegraphics[width=0.48\textwidth]{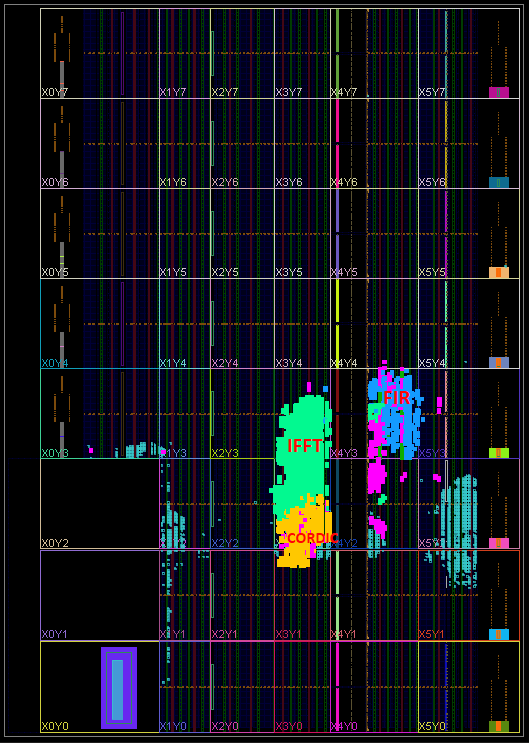}
    \caption{Area utilization of key synthesizer components in device logic cell view}
    \label{fig: UnilizationArea}
\end{figure}
\Cref{fig: UnilizationArea} shows the FPGA fabric area occupied by the implemented synthesizer. The entire grid represents the available FPGA fabric area.
The green, orange, and blue areas represent the IFFT, CORDIC, and M-path FIR of the synthesizer, respectively. The magenta areas are miscellaneous parts of the synthesizer, distinct from the three main components. The cyan areas primarily include DACs and other peripherals not part of the synthesizer.

\section{Conclusion}
The designed and implemented bias tone synthesizer, featuring an overlap-channel polyphase synthesis filter bank (OC-PSB) at its core, successfully lifts the bandwidth limitations of previous LUT-based approaches. Although the current BW is constrained to 256 MHz by the FPGA clock rate, parallel processing by 4 will give an effective clock rate of 1.024 GHz. Parallel processing only increases FPGA utilization without any changes in the designed sequence of operations. Results show that the synthesizer's performance meets essential readout requirements such as frequency resolution, number of probe tones, and SNR, while also providing enhanced readout capabilities like real-time (less than 20 µs) tone editing. Furthermore, it efficiently utilizes FPGA resources. The design process provides a different if not fresh view on the potential of PSB in its flexibility and efficiency as a multi-rate DSP technique. This effectively validates the feasibility of integrating PSB within arbitrary FPGA-KIDs-based telescope systems to achieve next-generation readout capabilities. Although it has some limitations, such as one tone per channel, this approach offers a versatile solution to the growing detector count and the concomitant challenges associated with the rapid advancements in MKID technology.

\subsection*{Future Work}
The implemented synthesizer currently achieves a quarter of the desired bandwidth pursued by the 850-GHz module, limited by the FPGA clock rate. However, future work involves extending the bandwidth to the target level through parallel processing within the FPGA fabric to effectively increase the clock rate, if directly over-clocking the FPGA is not feasible. This parallel-by-4 design will require up to four times the FPGA resources but remains primarily an engineering challenge, as the underlying algorithm remains unchanged.

Loop-back testing with the existing readout analysis channelizer, both with and without prototype detector arrays, will provide further performance metrics of the synthesizer as part of the readout system.

Additionally, further firmware and software integration between the synthesizer design and the Prime-Cam instrument readout system is necessary for practical operation on the Prime-Cam. Simplifying the control of the synthesizer can be achieved by implementing software masking for the tone editing process.

With the flexibility and scalability of the OC-PSB, further interesting testing and developments are anticipated.

\acknowledgments 
The CCAT-prime project, FYST and Prime-Cam instrument have been supported by generous contributions from the Fred M. Young, Jr. Charitable Trust, Cornell University, and the Canada Foundation for Innovation and the Provinces of Ontario, Alberta, and British Columbia. The construction of the FYST telescope was supported by the Gro{\ss}ger{\"a}te-Programm of the German Science Foundation (Deutsche Forschungsgemeinschaft, DFG) under grant INST 216/733-1 FUGG, as well as funding from Universit{\"a}t zu K{\"o}ln, Universit{\"a}t Bonn and the Max Planck Institut f{\"u}r Astrophysik, Garching.

\bibliography{main} 
\bibliographystyle{spiebib} 

\end{document}